\documentstyle[aps]{revtex}

\begin{document}

\title{Quantizing Constrained Systems: New Perspectives }
\author{L. Kaplan \\ Harvard Society of Fellows and
Department of Physics\\
Harvard University, Cambridge, MA 02138\\
N. T. Maitra\\
and\\
E. J. Heller \\
 Department of
Physics and Harvard-Smithsonian Center for
Astrophysics\\ Harvard University, Cambridge, MA 02138}
\maketitle

\begin{abstract} We consider  quantum mechanics on constrained
surfaces which have non-Euclidean metrics and variable Gaussian curvature.
The old controversy about the ambiguities involving
 terms in the Hamiltonian of order $\hbar^2$ multiplying the Gaussian curvature
is addressed.  We set out to clarify the matter by considering constraints
to be the limits of large restoring forces as the   constraint coordinates
deviate from their constrained values. We find additional ambiguous terms
of order $\hbar^2$ involving freedom in the constraining potentials,
demonstrating that the classical constrained Hamiltonian
or Lagrangian  cannot uniquely specify the quantization: the ambiguity
of directly quantizing a constrained system is inherently unresolvable.
However,   there is never any problem with a physical quantum system, which
cannot have infinite constraint forces and always fluctuates around the mean
constraint values.  The issue is addressed from the perspectives of adiabatic
approximations in quantum mechanics, Feynman path integrals, and
semiclassically in terms of adiabatic actions.
\end{abstract}

\section{Introduction}
\label{intro}

The controversy over the proper way to quantize
a constrained dynamical system has had a
long and interesting history in twentieth century physics.
One treatment of the subject opens by saying that ``if
you like excitement, conflict, and controversy, especially when
nothing very serious is at stake, then you will love the history of
quantization on curved spaces." \cite{schulman}
The possibility of resolving the ambiguities experimentally for rigid body
systems was the subject of a conference discussion in 1957 among
DeWitt, Wheeler, and Feynman \cite{schulman}. The problem
was also treated in Dirac's famous text of 1967 \cite{dirac}.
Yet many years later we still seem far from attaining a consensus
regarding quantization on curved spaces. The ``quasipermanent
discussion among the specialists" \cite{gutzwiller} has by no means
come to an end. In fact, a recent paper \cite{ali} laments that
``in spite of all the successes of quantum mechanics and after years of
efforts on quantization, we are still not absolutely sure about the correct
quantization of as simple a system as the double pendulum."
The subject also received early attention in the works of Cheng\cite{cheng}
and DeWitt\cite{dewitt}. In this paper we do not attempt to review the
extensive research on the subject of constrained quantization, 
and refer the interested reader to the literature\cite{morerefs}.
Our approach to this 
problem is quite independent of all previous methods.

The canonical quantization prescription has served us well in
defining a straightforward and unambiguous way of quantizing
a classical system in flat space. However, it immediately runs into
problems where constrained dynamics is involved. It is easy enough
to restrict the potential term to the constraint surface -- the
difficulty arises in treating the operator ordering ambiguities in
quantizing the kinetic term
$K=-{\hbar^2 \over 2 m}{1\over\sqrt g }{\partial\over\partial q^i}
(g^{ij}\sqrt g {\partial\over\partial q^j})$. These ambiguities, at order
$\hbar^2$, are proportional to the local Gaussian curvature of the
constraint surface, which is the only available
coordinate-invariant quantity. Thus for constrained systems with
flat constraints ({\it e.g.} the simple pendulum or a particle confined to
a one-dimensional curve), the ambiguity does not arise,
whereas for systems with constant curvature ({\it e.g.} motion on a sphere)
it leads only to a physically unobservable shift in the zero-point
energy. In the latter case, the semiclassical Van Vleck approximation
can be used to fix the ``correct" energy offset \cite{gutzwiller}. In three
dimensions this requires adding an $R$ term to the Laplacian
($R$ being the scalar curvature). For a two dimensional surface
of constant negative curvature, the
semiclassical approximation becomes exact (though only
at distances large compared to
the curvature radius)
if the quantity added to the Laplacian is $R/4$. Even if one
(somewhat arbitrarily?) requires
that Van Vleck should be as close as possible to being exact for the
constant curvature case, it is by no means clear that this prescription
is valid in the generic case of varying curvature, where the
ambiguities are most physically relevant.

One might think that Feynman's path integral approach \cite{feynman} would help
to resolve these uncertainties. However, as remarked by Schulman in
this context, ``there is no free lunch." In the path integral method, the
ambiguity arises in choosing how to incorporate the metric into
the kernel and in deciding at what point in the infinitesimal
time interval to evaluate functions of the metric \cite{schulman,ali}.
Two commonly used kernels give answers that differ by an effective $R/6$
contribution to the Laplacian (and naturally enough both differ from the
canonical quantization result). These issues are discussed at some length
in Ali's recent paper \cite{ali}, and here we will not go into the details.

Given this history, we have decided to attack this old problem
using a somewhat different, and in our view, more physically
motivated approach. The  principle governing our view is that arriving at a
unique quantization  is   essential only in real systems.
That is, quantization should be {\it physically} unambiguous.
For example,  quantization of the  usual rigidly constrained classical
double pendulum is not a physical problem, because the
   requirement that the pendulum lengths remain exactly
constant is not realizable. To make the problem as close as possible to
physical, the constraints should be imposed through a limiting procedure,
fluctuations remaining  physically allowable along the way.
We call this the ``limit
quantization,'' as opposed to   direct quantizations which do not pass
through a
limiting procedure.  This procedure is carried out in Sec.~\ref{lev} using
a Hamiltonian formalism, and later in Sec.~\ref{neepa} from a path
integral point of view. The limit quantization embodies  the idea of classical
constraints arising as the limit of ever larger restoring forces normal to
the constraint
surface, an idea discussed in Arnol'd, for example~\cite{arnold} (see also
Lanczos\cite{lanczos}).  In this
way the
constraint coordinates  appear as part of the full dynamical system in the
usual
Euclidean   metric.  As the constraint forces are increased,  the
constrained degrees of
freedom  acquire a very high frequency, and their actions remain
adiabatically constant
as the slow variables move.  This however does not imply  that the fast
variables (which
we here call the $r-$variables) are {\it energetically} decoupled from the slow
($\theta-$)
variables.  We  discuss the classical and semiclassical issues of
constrained dynamics
further in Sec.~\ref{semi}, where from
still another perspective  we will see that
the direct quantization is intrinsically ambiguous at order $\hbar^2$:
different
limiting constraint potentials  give the {\it same} classical constrained
motion but {\it
different} quantum dynamics  at order $\hbar^2$.

The uncertainty principle requires that in any
real physical system there will be quantum fluctuations around the surface
of constraint. By examining the effects of these fluctuations in
the limit of ever tighter constraints, we find unavoidable
ambiguities in the quantum dynamics along the surface of constraint.
Our claim is that these ambiguities are an inherent part of any
reasonable quantization procedure, and need not be more bothersome
in principle than the Aharonov-Bohm effect, which shares the
property of identical classical dynamics giving rise to different
quantum dynamics.  As the Aharonov-Bohm effect illustrates,  there is no good
reason why a quantum system should be uniquely determined by its classical
(or even semiclassical) limit, and the richness of the new phenomena
that can occur in physical quantum
systems can well be viewed as a positive aspect
of the quantum theory rather than as a shortcoming.

In any case,
we hope our contribution to the literature will be a valuable one, in
spite of this concluding warning by Schulman:
``Additional papers on curved space path integration are legion, and we
do not attempt to list them all. Some are correct; some are less correct.
Some have original features; some are less rich in this praiseworthy
property." \cite{schulman}

\section{Quantum Mechanics of Constrained Systems}
\label{lev}

A physically natural way of defining a constrained dynamical
system in classical mechanics is through a limiting procedure
where a strong attractive potential forces the system at any fixed energy
to live closer and closer to the constraint surface \cite{arnold}.
The effective classical dynamics obtained in this limit turns
out to depend only on the intrinsic properties of the constraint
surface ({\it e.g.} the curvature) and not on the details of the
constraining potential chosen on the embedding space. We will use an
analogous procedure in quantum mechanics and consider the
quantization of a constrained classical system as the limit
of quantizations of classical systems with large potentials
away from the surface of constraint. This approach is in accord
with our intuition about physical constraints. It is also
appealing because once a specific squeezing potential is chosen, the
resulting quantum mechanics is entirely unambiguous, and can
be obtained equivalently via a canonical or path integral method.
Ambiguities will nonetheless arise
in this approach, but are due to the freedom in selecting a constraining
potential.  The
effective quantum dynamics in the squeezing limit will have
corrections at order $\hbar^2$ which depend on the details
of the way in which this limit is taken. These ambiguities are
analogous to operator-ordering effects in canonical quantization
approaches, and to ambiguities which arise in the choice of
kernel (particularly in the choice of evaluation
points for functions of the metric) in path integral methods.
However, the physical ambiguities we find are considerably
more general in that they need not be functions only of the
intrinsic properties of the constraint space. Even for
a flat constraint space a large set of possible quantizations
are available, each having its own physical realization in an embedding space,
but all having the same classical and semiclassical
limits.  The existence of such quantum ambiguities in constrained systems
all leading
to the same classical physics in our view overshadows any attempt to resolve
the operator ordering questions and terms involving the curvature.

We begin with a Lagrangian
\begin{equation}
L={1\over 2}g_{ij}\dot{q}^i\dot{q}^j-V(q)-\tilde{V}_{\lambda}(q)
\label{eq:lag}
\end{equation}
on the full coordinate space, with flat metric $g_{ij}$ and
potential $V$ (we take the mass to be unity throughout).
The additional potential $\tilde{V}_{\lambda}$
enforces the constraint in the limit $\lambda\rightarrow\infty$.
Classically we require $\tilde{V}_{\lambda}$ to have the same
value everywhere on the constraint surface $S$, for each $\lambda$.
We now transform to the Hamiltonian
\begin{equation}
H={1\over 2}g^{ij}p_ip_j+V(q)+\tilde{V}_{\lambda}(q)\,,
\label{eq:hamilt}
\end{equation}
where $p_i=\partial L/\partial \dot{q}^i = g_{ij}\dot{q}^j$
is the momentum coordinate and $g^{ij}$ is the inverse of
$g_{ij}$. Canonical quantization now requires that we replace
$g^{ij}p_ip_j= \vec{p}\cdot\vec{p}$ with the operator
$-\hbar^2\nabla^2$, which in coordinate system $q$ has the form
\begin{equation}
-\hbar^2\nabla^2 = -\hbar^2 {1\over\sqrt g }{\partial\over\partial q^i}
(g^{ij}\sqrt g {\partial\over\partial q^j}) \,,
\end{equation}
where $g$ is the determinant of the metric ({\it i.e.} the
square of the volume element).
Notice that there is no operator ordering ambiguity here because
the full metric $g_{ij}$ is flat.

Now locally, near some region of the constraint surface $S$ we
may separate the coordinates $q$ into ``slow coordinates" $\theta$
parametrizing $S$ and ``fast coordinates" $r$ orthogonal
to the surface (see Fig.~(\ref{adia2})).
We will be looking for an effective theory of the
coordinates $\theta$ when the constraining potential
$\tilde{V}_{\lambda}^\theta(r)$ allows only small fluctuations in
$r$ (small compared to the length scale associated with the
physical potential $V$, the curvature scale of the constraint
surface $S$, and the scale on which $\tilde{V}_{\lambda}^\theta$
varies as a function of $\theta$). The quantum
wavefunction on the full space can be written  as
\begin{equation}
\Psi(q)=\Psi(\theta,r)=\phi_0(\theta)\Psi_{\rm GR}^{(\theta)}(r)+
\phi_1(\theta)\Psi_1^{(\theta)}(r)+\cdots\,,
\end{equation}
where $\phi_0$, $\phi_1, \ldots$ are arbitrary functions of the slow variables.
$\Psi_{\rm GR}^{(\theta)}(r)$ is the ground state of the fast
variables $r$, treating $\theta$ as a fixed parameter,
$\Psi_1^{(\theta)}(r)$ is the first excited state, and so on.
More precisely, $\Psi_{\rm GR}$, $\Psi_1$, etc. are the eigenstates
of the fast Hamiltonian
\begin{equation}
H_r = -{\hbar^2 \over 2} {1\over\sqrt g }{\partial\over\partial r^i}
(g^{ij}\sqrt g {\partial\over\partial r^j})+
\tilde{V}_{\lambda}^\theta(r)\,,
\end{equation}
where $i$ and $j$ are summed over the fast variables only and the
metric is evaluated at a fixed value of $\theta$. Note that $g$
here is the determinant of the full metric on the $(r,\theta)$
space.

Now we want to consider the effective Hamiltonian $H_{\rm eff}$
acting on the slow wavefunction $\phi_0(\theta)$ which multiplies
the ground state of the $r$ variables. We will argue
towards the end of this analysis that this effective theory is in fact
unitary in the limit $\lambda\rightarrow\infty$, as the transition
probabilities connecting the ground states to the excited states
of the fast variables disappear in the constraint limit (due
to adiabaticity).

We have mentioned previously that in order to obtain the correct
classical motion in the constraint limit, we must make
$\tilde{V}_{\lambda}^\theta(r=0)$ be independent of $\theta$ and
have the potential increase away from $r=0$. In order for
the quantum mechanics to have a sensible semiclassical limit,
we must further enforce that the energy of the ground state
$E_{\rm GR}^{(\theta)}$, defined by
\begin{equation}
H_r\Psi_{\rm GR}^{(\theta)}=
E_{\rm GR}^{(\theta)}\Psi_{\rm GR}^{(\theta)}\,,
\end{equation}
must be $\theta-$independent,
for any $\lambda$, and the energy of the first excited state
must be well separated from it, with the separation growing with
$\lambda$. This can be implemented in a number of ways, but two particularly
simple scenarios present themselves. In the first we take
$\tilde{V}_{\lambda}(q)=\lambda v(q)$, where $v$ is a smooth function
vanishing on the constraint surface and having its minimum there (but with a
non-singular second derivative matrix with respect to the fast variables).
Near the
constraint surface, the constraining potential will have the harmonic
oscillator form
\begin{equation}
\tilde{V}_{\lambda}^\theta(r)=r^iA^{(\theta)}_{ij}r^j+O(r^3)\,,
\end{equation}
where the matrix function $A^{(\theta)}$ is smooth. The ground states
$\Psi_{\rm GR}^{(\theta)}$ will to first approximation be harmonic
oscillator ground states with spatial
extent of order $\hbar^{1/2}/\lambda^{1/4}$
and energy $E_{\rm GR}$ of order $\hbar \lambda^{1/2}$. As discussed above,
the coefficient of this energy must be $\theta$-independent, to prevent
infinite effective forces in $\theta$ from arising
in the $\lambda\rightarrow\infty$ limit. The matrices $A^{(\theta)}$
can easily be adjusted to satisfy this condition. In the case of more
than one constraint variable, this still allows for much freedom in
the function $A^{(\theta)}$ and the resulting eigenstates
$\Psi_{\rm GR}^{(\theta)}$ (energy can flow from one fast degree of
freedom to another as a function of $\theta$, as long as the total
energy remains constant). One may choose to adjust (some of)
the anharmonic parts
of the potential as well but this is not really necessary or natural. The
quartic term will lead to order $\hbar^2$ ($\lambda$-independent)
corrections to the ground state energy and thus (if these
anharmonic corrections
vary with $\theta$ which in general they will) to the effective
potential in $\theta$. \footnote{For the purpose of dimension counting,
every power of r in an expression contributes a factor of
$\hbar^{1/2}/\lambda^{1/4}$ to the expectation value whereas
$\partial/\partial r$ acting on $\Psi_{\rm GR}$ (but not the metric)
 contributes
$\lambda^{1/4}/\hbar^{1/2}$. Derivatives acting on the ``slow"
wavefunction $\phi_0(\theta)$ and the metric $g$, as well as
$\partial/\partial\theta$ acting on $\Psi_{\rm GR}$ all produce factors
of order unity.}
We will later see that this effect is of the
same order as other terms that are encountered in the effective
potential. We might also mention at this point that if in finding the
ground state of $r$ at a given $\theta$ we used normal
coordinates centered at $(\theta,r=0)$,
$g_{ij}=\delta_{ij}+O(r^2)$ for fixed $\theta$,
and neglected the $O(r^2)$ curvature
correction, we would also make an error of order $\hbar^2$ in the ground
state energies and therefore in the effective potential for $\theta$.

A second scenario involves a hard-wall potential with the cross-section
of the allowed region shrinking towards $S$ as $\lambda\rightarrow\infty$.
Thus we may instead take
\begin{eqnarray}
\tilde{V}_{\lambda}^\theta(r)= & 0, \; a^{(\theta)}(\lambda r) < 1 \\
& \infty, \; a^{(\theta)}(\lambda r) > 1\,,
\end{eqnarray}
where the function $a^{(\theta)}$ defines a $\theta-$dependent region
shrinking as $1/\lambda$ in which the fast eigenstates live. The
energy of the ground state is of order  $\hbar^2\lambda^2$ and can again
be adjusted to be $\theta-$independent while allowing for substantial
freedom in the shapes of the ground states. In the following we will
focus on the original (smooth potential) scenario,
but this one could be used to obtain
similar results.

We now see what happens when we compute the effective Hamiltonian for
the slow variables $H_{\rm eff}(\theta,\partial/\partial\theta)$, defined
by
\begin{equation}
\langle\phi'_0|H_{\rm eff}|\phi_0\rangle \,=\, \langle\Psi'|H|\Psi\rangle
\end{equation}
with
\begin{equation}
\Psi(\theta,r)=\phi_0(\theta)\Psi_{\rm GR}^{(\theta)}(r)\,.
\end{equation}

The potential $V(q)$ clearly induces an effective potential
$V(\theta,r=0)$, up to corrections which disappear in the
constraint limit, since $V(\theta,r)$ varies slowly in $r$ over
the extent of $\Psi_{\rm GR}^{(\theta)}(r)$. The constraint potential
$\tilde{V}_\lambda$, combined with the fast part of the kinetic term,
gives, by assumption, a $\theta$-independent effective potential
which we can therefore drop. We are left with the slow $(\theta)$ part
of the full $q-$space Laplacian, as well as mixed terms involving one
derivative each with respect to the slow and fast variables.

In the Born-Oppenheimer spirit, consider first the action of the ``slow"
part of the Laplacian on the
full wavefunction. We are interested in matrix elements of the form
\begin{equation}
\langle\phi_0'\Psi_{\rm GR}|
-{\hbar^2 \over 2} {1\over\sqrt g }{\partial\over\partial \theta^i}
(g^{ij}\sqrt g {\partial\over\partial \theta^j})
|\phi_0 \Psi_{\rm GR}\rangle \,.
\end{equation}
The two derivatives can act on the slow wavefunction $\phi_0$, the
ground state $\Psi_{\rm GR}$, or on the metric and its determinant.
Consider terms where both derivatives act on the ground state or the
metric. After
integrating over the fast variables, this leads to
$\langle\phi_0'|-\hbar^2 f(\theta)|\phi_0\rangle$, an effective potential
scaling
as $\hbar^2$. \footnote{The derivatives acting on the metric only will give
the usual curvature-dependent corrections to the effective potential.
Terms coming from derivatives acting on the ground state
are of the same order and will give rise to additional corrections.}
(Note that both the metric and the ground state are
changing over typical scales of order unity in $\theta$.) Furthermore,
this effective potential is $\lambda-$independent for large $\lambda$,
since $\Psi_{\rm GR}$ maintains its functional form (that of a gaussian
plus higher order corrections) and simply
shrinks towards $r=0$ as $\lambda\rightarrow\infty$.

Now suppose one of the derivatives acts on the metric or on $\Psi_{\rm GR}$,
while the other acts on the slow wavefunction. We then obtain
\begin{equation}
\langle\phi_0'|-\hbar^2 h^i(\theta){\partial\over\partial \theta^i} +
{\rm h.c.}|\psi_0\rangle\,,
\end{equation}
where $h^i$ are real functions of $\theta$ (because the ground state
$\Psi_{\rm GR}$ can be chosen to be real, and the metric is real as well).
Taking the hermitian part
we again obtain an effective potential of order $\hbar^2$. Finally,
if both derivatives act on the slow wavefunction $\phi_0$, we obtain
the usual kinetic term for the slow variables, {\it including} order $\hbar^2$
terms proportional to the curvature of the constraining surface,
as discussed in the preceding section. This concludes the discussion of the
pure $\theta-$derivative part of the kinetic energy, and we move on to
the mixed terms next.

 If the $\partial/\partial r$
derivative acts on the metric, we obtain a contribution of order
$\hbar^2$ to the effective potential as before (all derivatives are
of order unity). If $\partial/\partial r$ acts on $\Psi_{\rm GR}$
while $\partial/\partial \theta$ acts on $g$ or the wavefunction
$\phi_0$, we obtain
\begin{equation}
\hbar^2 \langle\Psi_{\rm GR}|b^{\theta}(r){\partial\over\partial r} + {\rm
h.c.}
|\Psi_{\rm GR}\rangle\,,
\end{equation}
where $b$ is a slowly varying real function. So to this order in $\hbar$
only the commutator of $b$ with $\partial/\partial r$ survives, and once
again the resulting contribution to the effective potential is of
order $\hbar^2$. The remaining term is the one where both derivatives act
on the ground state; it has the form
\begin{equation}
\hbar^2 \langle\Psi_{\rm GR}|c(\theta,r){\partial\over\partial \theta}
{\partial\over\partial r} |\Psi_{\rm GR}\rangle\,.
\end{equation}
Since $\Psi_{\rm GR}(r)$ is even in $r$, the slowly varying function
$c(\theta,r)$ must be
expanded to first order in $r$ to get a non-vanishing contribution. We
then obtain yet another contribution to the effective potential of order
$\hbar^2$ (since $r {\partial\over\partial r}$ acting on $\Psi_{\rm GR}(r)$
is of order unity).

Finally, we argue by adiabaticity that there is no mixing between
$\phi_0(\theta)$ and the
excited wavefunctions $\phi_1(\theta)$, etc. This follows because
the splitting between $\Psi_{\rm GR}^{(\theta)}$ and the excited states of
$r$ scales as $\hbar\sqrt\lambda$ in the constraint limit (or
as $\hbar^2\lambda^2$ in the case of a hard wall potential), whereas
the matrix elements do not grow with $\lambda$. So assuming a finite
amount of energy in the initial $\phi_0(\theta)$ wavefunction,
transition probabilities vanish in the $\lambda\rightarrow\infty$
limit, and we obtain a probability-conserving effective quantum
mechanics. The effective Hamiltonian is
\begin{equation}
H_{\rm eff}= -{\hbar^2 \over 2}
{1\over\sqrt g }{\partial\over\partial \theta^i}
(g^{ij}\sqrt g {\partial\over\partial \theta^j})+V(\theta)+
\hbar^2V_{\rm eff}(\theta)\,.
\end{equation}
This is the key result of this paper. $V_{\rm eff}$ is essentially an
arbitrary smooth function, depending in a complicated way on both the
curvature properties of $S$ and on the constraining potential
around $S$ (together these determine the ground state wavefunctions).
$\sqrt g $ can be taken to be the square root of the determinant of
either the full metric or of the metric restricted to the constraint
surface. Any difference between these expressions can be absorbed
into $V_{\rm eff}$.

\section{The Lagrangian Formalism}
\label{neepa}

A   path integral analysis also shows that
quantization on the constrained surface is ambiguous at order
$\hbar^2$.
We begin with the propagator on the full coordinate space.
Because the space is flat, this is
given by the
usual Feynman prescription $\int {\cal
D}(q(\tau))\exp(iS(q(\tau))/\hbar)$ where $S(q(\tau))= \int L(q(\tau),
\dot q(\tau), \tau) d\tau$. The Lagrangian $L$ on the full space is
given by Eq.~\ref{eq:lag} in Section~\ref{lev}; we follow the
notation introduced there.  We extract the effective dynamics for the
``slow'' variables $\theta$ by performing a trace over the
``fast'' variables $r$ in the adiabatic limit.
Specifically, we  calculate $\langle \theta_f
\vert\rho(t)\vert\theta_o\rangle$ where $\rho(t)= {\rm Tr}_r(W(t))$ is
the trace over the fast variables of the density operator on the full
space. The time $t$ is taken to be small on the time scales of
$\theta$-motion but $r$ may undergo many oscillations during this time.
Noting that
the evolution of the full density operator is given by
$W(t)=e^{-iHt/\hbar}W(0)e^{iHt/\hbar}$, and inserting identities in the
form of complete sets of states, we have
\begin{equation}
\langle \theta_f\vert\rho(t)\vert\theta_o\rangle =
\int dR dr' dr'' d\theta' d\theta'' \langle \theta_f R \vert
e^{-iHt/\hbar}\vert\theta' r'\rangle \langle \theta' r'\vert
W(0)\vert\theta'' r''\rangle \langle \theta'' r''\vert
e^{iHt/\hbar}\vert\theta_o R\rangle\,.
\end{equation}
We take the initial density operator to be the ``product'' state $\vert
\Psi_{\rm GR}^{(\theta)}\rangle \otimes \vert \phi \rangle$, where
the initial state of the fast variables is the (purely real) ground state
$\Psi_{\rm GR}^{(\theta)}(r)$, with $\theta$ regarded
as a fixed parameter (see Section~\ref{lev}). $\phi (\theta)$
is an arbitrary initial wavefunction of the slow variables $\theta$.
Replacing the two propagators by path integrals, we then have
\begin{equation}
\langle \theta_f\vert\rho(t)\vert\theta_o\rangle = \int d\theta'
d\theta''\int_{\theta (0)=\theta'}^{\theta(t)=\theta_f}{\cal
D}\theta(\tau)\int_{\tilde\theta
(0)=\theta''}^{\tilde\theta(t)=\theta_o}{\cal
D}\tilde\theta(\tau)\phi(\theta')\phi^*(\theta'') {\cal
F}\left(\theta(\tau),\tilde\theta(\tau)\right)
\label{eq:rhoF}\,,
\end{equation}
where
\begin{eqnarray}
\nonumber
{\cal F}\left(\theta(\tau),\tilde\theta(\tau)\right)&=&\int dR dr' dr''\\
\nonumber
&\times&\int_{r(0)=r'}^{r(t)=R}{\cal D}r(\tau)\int_{\tilde
r(0)=r''}^{\tilde r(t)=R}{\cal D}\tilde r(\tau)e^{iS^{(1)}/\hbar
-iS^{(2)}/\hbar}\Psi_{\rm GR}^{(\theta'')}(r'')\Psi_{\rm GR}^{(\theta')}(r')
\,.\\
\end{eqnarray}
$S^{(1)}=S(r(\tau);\theta(\tau))$ and $ S^{(2)}=S(\tilde
r(\tau);\tilde\theta(\tau))$ are the actions along the respective
paths.  In performing the path integrals over the fast variables,
$\theta(\tau)$ ($\tilde\theta(\tau)$) is to be treated as an
external forcing function with the property $\theta(0)=\theta',\,
\theta(t)=\theta_f$ (resp. $\tilde\theta(0)=\theta'',\,
\tilde\theta(t)=\theta_o$).

We evaluate ${\cal F}$ using stationary phase with respect to
$\hbar$ on the path integrals, trace, and the integrals over the
intermediate fast variables.  In fact the stationary phase approximation
becomes exact in the
adiabatic limit: we shall now show that the errors due to stationary phase
evaluation
are of order $\hbar^{1/2}/\lambda^{1/4}$, so they vanish as
$\lambda\to\infty$.
Recall that
\begin{equation}
\int g(x)e^{if(x)/\hbar}dx \approx_{\rm sp}\sum\sqrt{\frac{2\pi
i\hbar}{f''(x_n)}}
g(x_n)e^{if(x_n)/\hbar}\left(1+O(\hbar^{1/2}\frac{f'''(x_n)}{f''(x_n)^{3/2}})\right)\,,
\end{equation}
where the sum is over stationary phase points $x_n$ satisfying
$f'(x_n)= 0$.  Stationary phase evaluation of path integrals, although
different in the details, scales in the same way. $f(x)$ corresponds
to the action $S$. The stationary paths are the classical paths and we
now argue that $S_{\rm cl}\sim O(\lambda^{1/2})$ to leading order in
$\lambda$. The leading behavior of the action in
the fast variables arises from the kinetic term in $r$ and the
 constraint potential: $L_{r}\sim g_{ij}\dot r^i\dot r^j/2 - \tilde
V_{\lambda}^{\theta}(r)$, where the sum is over fast variables only. This
Lagrangian
corresponds to the fast Hamiltonian of Eq.~\ref{eq:hamilt} discussed in
Section~\ref{lev}, which gives harmonic motion for $r$ at least
throughout the range of the fast variable ground state. Because the
initial fast state is this ground state, the leading behavior of
$e^{iS/\hbar}$ gives
\begin{equation}
\int dr'\int {\cal D}r(\tau)
e^{iS^{(1)}/\hbar}\Psi_{\rm GR}^{(\theta')}(r')\sim
e^{-iE_{GR}t/\hbar}\Psi_{\rm GR}^{(\theta')}(r)\,,
\end{equation}
where we note that $E_{\rm GR}\sim O(\hbar\sqrt{\lambda})$ is required to be
$\theta$-independent to avoid infinite torques acting on the slow
variables (Section~\ref{lev}).  This shows the leading  behavior of the
action on the ground state is simply to multiply it by an evolving
phase. The fast variable $r(t)$ does not stray from  the domain of faithful
harmonic approximation to the constraint potential. Subleading terms
 are a factor of $1/\sqrt{\lambda}$ smaller. Thus the
dominant behavior of the action and its derivatives is simply that of a
harmonic potential of frequency $O(\sqrt{\lambda})$.  The errors in
the stationary phase evaluation of the path integrals are of order
$\hbar^{1/2}/\lambda^{1/4}$, vanishing in the adiabatic limit.  The
errors in using stationary phase to compute the trace and integrals
over the intermediate variables $r'$ and $r''$ scale in the same way.
In the harmonic approximation to the action and
the ground state it is readily seen that the stationary phase points are
$R=r'=r''=0$, the constrained value of the fast variable.
Corrections to this approximation due to
subleading terms vanish in the adiabatic limit, $\lambda \to\infty$.

We have then,
\begin{eqnarray}
\nonumber
{\cal F}(\theta(\tau), \tilde\theta(\tau))&=&\left(\frac{\partial^2
\ln\Psi_{\rm GR}^{(\theta')}}{\partial r'^2}\right )^{-1/2}\left
(\frac{\partial^2 \ln\Psi_{\rm GR}^{(\theta'')}}{\partial
r''^2}\right)^{-1/2}\\
&\times&\left\vert\frac{\partial^2S^{(1)}_{\rm cl}}{\partial R\partial
r'}\right\vert^{1/2}\left\vert\frac{\partial^2S^{(2)}_{\rm cl}}{\partial
R\partial r''}\right\vert^{1/2}e^{iS^{(1)}_{\rm cl}/\hbar -
iS^{(2)}_{\rm cl}/\hbar}\,,
\end{eqnarray}
where the derivatives are evaluated at $R=r'=r''=0$.
$S^{(1)}_{\rm cl}$(resp. $S^{(2)}_{\rm cl})$ is the action along the
classical path starting and ending at the constraint value $r=0$
subject to the Lagrangian of Eq.~\ref{eq:lag} with
$\theta(\tau)$ (resp. $\tilde\theta(\tau)$) treated as an undetermined
forcing function.  In the adiabatic limit, we shall now show that the
action exponent is
just the action for the reduced slow variable system on the
constraint surface, {\it i.e.} the action we would have written down had we
 begun in
the reduced space. The kinetic
term in $r$ together with the constraint potential give a
$\theta$-independent term in the exponent as discussed above.
This, being a constant energy shift as far
as $\theta$ is concerned, can be neglected.  The mixed kinetic term
goes to zero: we may expand $g_{ij}(r,\theta)$ about $r=0$ and take
functions of $\theta$ out of the time-integral as they are slowly
varying functions of time:
\begin{eqnarray}
\nonumber
\int g_{ij}(r,\theta)\dot r^{i} \dot \theta^{j} dt &=&
g_{ij}(0,\theta)\dot\theta^{j}\int_0^{t}\dot r^i dt + \frac{\partial
g_{ij}}{\partial r^p}\dot\theta^j\int_0^t \dot r^i r^p dt\\
&+&O(\hbar^{3/2}/\lambda^{1/4})\,,
\end{eqnarray}
where the derivatives of the metric are evaluated at $r=0$. The first term
on the right-hand side is zero, since the integral gives $r^i(t)-r^i(0) = 0$
due to the stationary phase condition on the endpoints. For $p=i$, the
 integral in
the second term is zero for the same reason (the integrand is $d
(r^{i2})/dt$). For $p\neq i$ the integral averages to zero, because different
directions
of the fast variables generically oscillate at different frequencies. The
higher order terms in the expansion vanish in the adiabatic limit (they
involve at least two powers of $r$ but only one power of $\dot r$ so scale at
least as $(\hbar/\lambda^{1/2})(\hbar^{1/2}\lambda^{1/4})$).
We are thus left with the kinetic term in the slow variables only, with
the metric evaluated at the constraint value for $r$.

Inserting ${\cal F}(\theta(\tau), \tilde\theta(\tau))$ into
Eq.~\ref{eq:rhoF} gives the reduced density matrix at time $t$.
This factorizes into a part involving ($\theta, \theta_f$, and
$\theta(\tau)$) and a part involving ($\theta',\theta_o,$ and
$\tilde\theta(\tau)$).
This implies that the reduced density matrix factorizes:
$\langle\theta_f\vert\rho(t)\vert\theta_o\rangle=
\langle\theta_f\vert\theta(t)\rangle\langle\theta(\tau)\vert\theta_o\rangle$,
{\it i.e.} the final $\theta$-state is pure and so we may describe it in terms
of a wavefunction. Adiabaticity has thus uncoupled the fast and slow degrees of
freedom. We have
\begin{equation}
\phi(\theta_f,t) = \int d\theta\int {\cal D}\theta(\tau)
 A(\theta(\tau))e^{iS/\hbar} \phi(\theta,0)\,,
\label{eq:prop}
\end{equation}
where
\begin{equation}
A(\theta(\tau)) = \left(\frac{\partial^2 {\rm {\rm ln}}
\Psi_{\rm GR}^{(\theta)}}{\partial
r^2}\vert_{r=0}\right)^{-1/2} \left\vert\frac{\partial^2
S^{(1)}_{\rm cl}}{\partial R \partial r}\vert_{R=r=0}\right\vert^{1/2}\,.
\end{equation}
$S=S^{(1)}_{\rm cl}$ is the action function for the $\theta$ variable on the
constrained surface: $S=\int L dt$ where the Lagrangian is $L=
\frac{1}{2}g_{ij}\dot \theta^i\dot\theta^j - V(\theta)$, with $g_{ij}$
now the metric on the curved space defined by the constraint surface and
the sum is over the slow ($\theta$-)variables only.

We may compare this to the expressions discussed in Ali, where one
works in the reduced space from the start. Different Feynman kernels
are postulated to attempt to account for the curvature of the
constrained surface \cite{schulman,ali,feynman}. There, for infinitesimal
time $t$,
\begin{equation}
\phi(\theta_f,t)=\int d\Omega(\theta) \tilde
G(\theta,\theta_f)e^{iS/\hbar}\phi(\theta,0)\,.
\end{equation}
$d\Omega(\theta)=\sqrt{g(\theta)}d\theta$ is the volume element at
$\theta$.  Candidates for $\tilde G(\theta,\theta_f)$ which are often
considered in the literature include the identity operator and
$g(\theta)^{-1/4}D^{1/2}g(\theta_f)^{1/4}$, where $g$ is the
determinant of the metric on the curved space, and $D$ is the
Van-Vleck determinant ($D=\det(-\partial^2
S/\partial\theta\partial\theta_f)$).
The different
choices give rise to Hamiltonians which differ at order $\hbar^2$ by a
certain fraction of $\hbar^2 R$. Ali demonstrates this by
choosing locally normal coordinates $\xi$ to evaluate
the integrals and considering
infinitesimal time  so that the exponent and prefactor can be expanded in
 $\xi$.
The action exponent is then at lowest order (in $\xi$) quadratic,
and the resulting Gaussian integrals are readily performed. Only the
even order terms in the expansion of the prefactor about the initial
value $\theta$ contribute. The $t\to 0$ limit is taken and an effective
Hamiltonian can be extracted from the resulting differential
equation. The expansion of the potential exponent and the
 initial wavefunction give the Hamiltonian $-\hbar^2\nabla^2/2m +V$,
and it is the quadratic term in the
expansion of $\sqrt{g(\xi)}\tilde G(\xi)$ that gives rise to the
$\hbar^2 R$ discrepancies in the effective Hamiltonian. (Higher
order terms give corrections at higher order in $\hbar$.)

The key point is that  we may apply the same manipulations to our
Eq.~\ref{eq:prop},
where $A(\theta(\tau))$ plays the role of the prefactor
$\sqrt{g(\theta)}\tilde G(\theta)$. Transforming to locally normal
coordinates in which there are no linear terms in the action exponent, and
expanding $A(\theta(\tau))$ about $\theta$
leads to finite $\hbar^2$ corrections in the effective Hamiltonian
as in Ali's approach described above.  However, it is important to
recognize that our corrections emerge from reducing the full,
unambiguous    coordinate space evolution
down to the reduced curved space by taking the constraint limit,
 whereas Ali's corrections  arise from using a modified Feynman kernel for
 curved space.
We observe that our corrections depend not only on the curvature but
also on the constraint potential.  These conclusions are in accord with the
results of the previous section and provide  additional insight into the
ambiguity and connection with the literature.

\section{Classical and Semiclassical Perspective}
\label{semi}

Finally we provide a third, classical and semiclassical
 perspective on the ambiguities
inherent  in quantizing  constrained systems.  The classical  point of
view on constraints as limits of strong restoring
forces is instructive, and nicely complements the
quantum discussion. In classical mechanics, adiabaticity plays an important
role.  Adiabaticity of
classical actions also played  an important role in the old quantum theory.
Here we will see
that the semiclassical approach  confirms the inherent order $\hbar^2$ problem
of quantizing a constrained system. Again the plan is to
start with the system in the full Cartesian space, transform
to curvilinear coordinates and the Laplace-Beltrami kinetic
energy operator with no ambiguity, and begin imposing stiff force
constants about   fixed values of one or more of the coordinates.  The
result in the limit of infinite restoring
forces is constrained dynamics  on a Riemannian manifold.  Again
an ambiguity will arise because we  can arrive at identical classical
constrained dynamics in ways which differ quantum mechanically.

The idea that classical constraints are  the limit of merely stiff
degrees of freedom is a natural one, though not often discussed.
Arnol'd includes a brief treatment of this in his famous book in  a
chapter devoted to Lagrangian dynamics on manifolds~\cite{arnold}.
Arnol'd states that in the limit $\lambda\to \infty$ the classical
dynamics satisfies Lagrange's equations of motion
\begin{equation}
{d\over dt} \left ( {\partial L_* \over \partial \dot \theta_i}\right ) =
\left ( {\partial L_* \over \partial  \theta_i}\right )\,,
\end{equation}
where
\begin{equation}
L_*(\{\theta\},\{\dot \theta\}) = T\vert_{r_i = 0; \dot r_i = 0} -
V\vert_{r_i = 0}\,,
\end{equation}
where $T$ is the kinetic energy and $V$ any non-constraint potential energy
of the full system.
However there is an important difference for the present purposes:
we need to consider   very large energies stored in the   constrained
coordinates, corresponding to the quantum zero point energy
climbing to  arbitrarily high values as $\lambda\to \infty$.  This we can do,
however, using   the theorem on the adiabaticity of classical actions under
slow variation of a parameter.

The issues which confront the classical approach to constraints in the
high energy constraint limit are 1)  energy stored in the fast variables
generally varies as a function of the slow variables,
even though actions are constant, thus raising the spectre of large energy
exchanges with the slow degrees of freedom, 2) the possibility that the
slow coordinates are chaotic, and 3) the possibility that the constraint
coordinates can exchange energy among themselves, perhaps chaotically.

\subsection{One dimension}

In the simple case of one dimensional motion with an adiabatic change
of  one or more parameters, the variation of the action decreases exponentially
as the rate of variation of the parameters decreases~\cite{landau}.  However
as the phase space diagram in Fig.~(\ref{adia},a) shows,
the energy generally changes significantly
as the parameters change, as measured by the contemporary value of the
Hamiltonian.  For example, in the case of  a one dimensional harmonic
oscillator,
suppose we slowly adjust the frequency $\omega(t) = \lambda(t)\omega_0$
toward higher
values, beginning with some energy $E_0$ in the oscillator:
\begin{equation}
H = {1\over 2} p^2 + {1\over 2}\lambda(t)\omega_0^2 q^2.
\end{equation}
Since the action is constant, the area of the ellipse enclosed by the
trajectory is constant, but this means that at $q=0$ the momentum
must increase.  However at $q=0$, $H=p^2/2$, so the energy increases as
$\omega(t)$ increases adiabatically.  Specifically, with $\omega \propto
\sqrt \lambda$,
we have $p \propto \lambda^{1/4}$ and $E\propto \lambda^{1/2}$.

Now suppose we have only one constraint coordinate $r$.  The situation
is depicted in Fig.~(\ref{adia2}).
The slow
 $\theta-$coordinates act as adiabatic parameters.  Whatever their
motion, in the limit of large, fixed $\lambda$  the action of the
$r-$motion is preserved during the course of the $\theta$ motion.
Clearly if we are to avoid any of the large zero point energy of the
$r-$coordinates
finding its way   into the $\theta-$coordinates we must restrict the
confining potential $V(r;\theta_1, \theta_2, \ldots)$  to vary with
$\theta_1, \theta_2,
\ldots$ in a way that maintains fixed energy for fixed action. This condition
does not fix the potential uniquely, but  is
possible for a wide class of potentials, for example satisfying
\begin{equation}
\label{co1}
E_\lambda(I) = \lambda^{1/2}(c_1 I + c_2 I^2 +\cdots )\equiv \lambda^{1/2}
\epsilon(I)
\end{equation}
with
\begin{equation}
\epsilon(\hbar/2) = \epsilon_0
\label{co2}
\end{equation}
independent of $\lambda$.
The action $I$ is defined as usual as
\begin{equation}
I = {1\over 2 \pi} \oint p_r(E,\lambda)  \ dr\,.
\end{equation}
An example showing two potentials (or rather their phase space contours
at identical actions and energies)   satisfying
this  requirement  is shown in Fig.~(\ref{adia},b). The area enclosed by the
dashed and solid contours is the same and the momentum at $q=0$ (where
we assume $V=0$ for both potentials) is also
identical, showing the energy is the same
for the two potentials at that action.
Any potential satisfying Eqs.~(\ref{co1}) and (\ref{co2})
at fixed values of the $\theta-$coordinates is an allowed constraint
potential at that point in $\theta$ space; the independence of
Eqs.~(\ref{co1}) and (\ref{co2})
under changes in the $\theta-$coordinates must then be arranged to
ensure energy remaining fixed in the $r-$variable.

The {\it semiclassical} estimate of the energy in the $r-$coordinate
is thus independent of the $\theta-$coordinates, since that energy
is just given by the classical energy at action $I = \hbar/2$; this
is what we set constant by definition.  However it is not the semiclassical
energy we want, but rather the {\it exact quantum} energy. Now the
key
fact is that {\it the exact and semiclassical energy  differ in order}
$\hbar^2$.
This  follows from the derivation of the WKB energy equation, which is an
expansion in $\hbar$.  Thus, the   freedom to choose among potentials
which lead to  identical classical constrained motion as $\lambda\to\infty$
leads unavoidably to a quantum ambiguity at order $\hbar^2$, in agreement
with the discussion in the previous two sections.  This is  another
confirmation of the view forwarded there, and is the main result of this
section.

\subsection{Chaos in $\{\theta\}$ Variables}

The issue of chaos in the $\{\theta\}$ variables is easily disposed of:
it makes no difference, since in the limit $\lambda\to\infty$ the frequency
of the $r-$variable is far from reach of the slow $\theta-$coordinates,
which just act as slow parameter changes on the $r-$variable motion. The
fact that the changes of the $\theta-$coordinates are not quasiperiodic
in the chaotic case is of no consequence to the adiabaticity of the
$r-$coordinates.

\subsection{Resonance or Chaos in $\{r\}$ Variables}

Interaction between several $r-$variables can be avoided  by
independent control of as many $\lambda-$parameters as $r-$coordinates.
With this control we can force the frequency ratios
$\omega_i/\omega_{i+1}\to 0$
even as each  $\omega_i\to \infty$.  By this ruse we avoid any low order
resonance leading to energy  transfer between the $r-$variables.  The
concern that
high order resonances must always exist is softened by the   fact
that even if the external control parameters ($\theta-$variables) are
effectively fixed
 and the high order resonance is allowed to act, the energy exchange
(resonance width) is expected to be exponentially small in the
winding number (frequency ratio)  for smooth potentials, as it depends
on Fourier coefficients of the order of the winding number.  In effect,
good actions are maintained in the $r-$variables as $\lambda\to\infty$.

In the opposite extreme of full chaos in the $r-$variables, we appeal to
Liouville's theorem applied to
 the  invariance of the total  phase space volume enclosed
by the  energy surface in the $r-$motion in the limit that the
$\theta-$variables
are slow enough to be considered to be adiabatically
separate external parameters.  In addition, we need Hertz's theorem,
which states that in an ergodic system the energy shell maps onto
another energy shell in the adiabatic limit of changing
parameters~\cite{hertz,reinhardt}.  We
may write
$E = E_{\{\theta\}}({\cal V})$, where ${\cal V}$ is the phase space volume.
As long as the constraint potentials are tuned to preserve
this energy as a function of the $\theta-$variables, {\it i.e.}
$E_{\{\theta\}}({\cal V}) = E_{\{\theta'\}}({\cal V})$ for all ${\{\theta'\}}$
and fixed ${\cal V}$,
there will again be no forces on them from the constraints.
Having arranged the potentials in this way, there remains the question of the
semiclassical energy for the chaotic dynamics.  This can be given by periodic
orbit theory~\cite{po}.  Somewhat paradoxically, periodic
orbit theory has been shown to give excellent results for low energy
eigenvalues in a number of systems, breaking down at higher energies.  We need
 the lowest (zero point) energy, which at least formally has an error of  order
$\hbar^2$, so once again we reach the same conclusion about the order $\hbar^2$
ambiguity  in the quantization procedure.

\section{Conclusion}

We have shown from three different points of view that the ambiguities
which arise in the quantization of a constrained system are physical
in nature and cannot be resolved by finding the ``right'' mathematical
formalism. Considering the constrained surface as a reduced sub-space of
a larger system, we have found that terms of order $\hbar^2$ inevitably
arise in the effective description. These terms depend on the details of
the forcing potential as much as on the intrinsic geometry of the constrained
surface, but there is no ambiguity in the classical or even the semiclassical
dynamics. Thus we obtain a large set of equally valid ``quantizations'' of
the same classical system, each corresponding to a particular constraining
process. In any physical situation, one such constraining process is the
correct one, and the ambiguities disappear. In the absence of detailed
knowledge about the nature of the constraints, additional ``quantization
conditions'' must be selected before one can speak of quantizing any
classical dynamics on a curved space.

Our discussion has been cast in terms of holonomic constraints (which
can be written in terms of the vanishing of one or more functions of the
coordinates)
but the idea of reaching rigid constraints via a limiting process
applies also to nonholonomic constraints, which can be couched only
in terms of nonintegrable differentials.

\section{Acknowledgments}
This  research
was supported by the National Science Foundation under grant number
CHE-9321260, and through ITAMP at the Harvard-Smithsonian Center for
Astrophysics.  We thank Michael Efriomsky  and Sidney Coleman for
helpful comments.

\clearpage
{\bf Figure Captions} 
\begin{figure}[h]
\caption{Schematic showing the fast motion along $r$ and slow motion along
$\theta_1, \theta_2$ coordinates for finite confining forces (finite
$\lambda$).}
\label{adia2}
\end{figure}
\begin{figure}[h]
\caption{a. An area preserving (adiabatic) change of Hamiltonian
parameters generally changes the energy, as indicated by the increase in
$\vert p\vert$ at $q=0$, where the potential $V=0$. b)  {\it Some}
slow deformations
of the  Hamiltonian do leave the energy unchanged.}
\label{adia}
\end{figure}

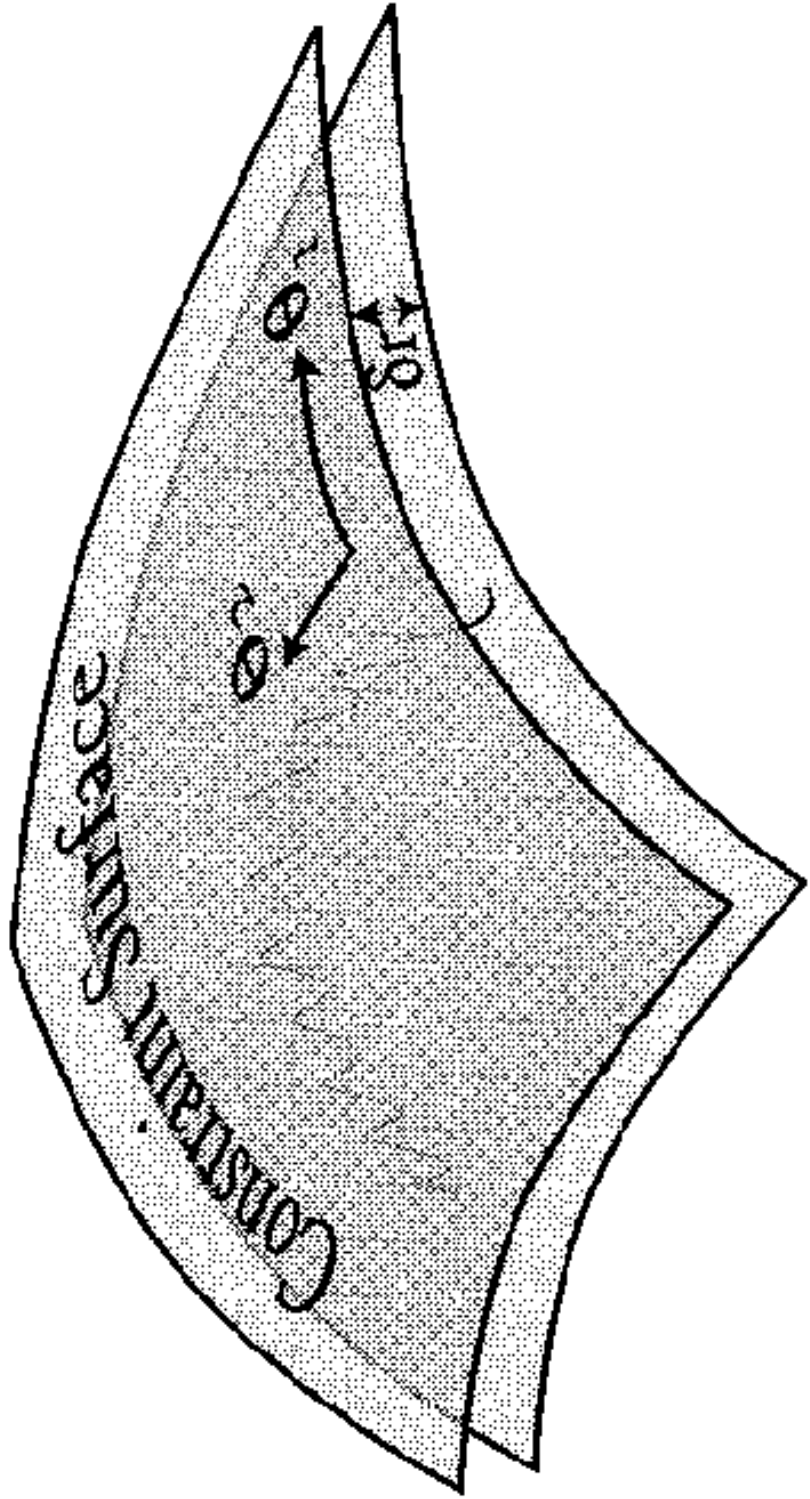
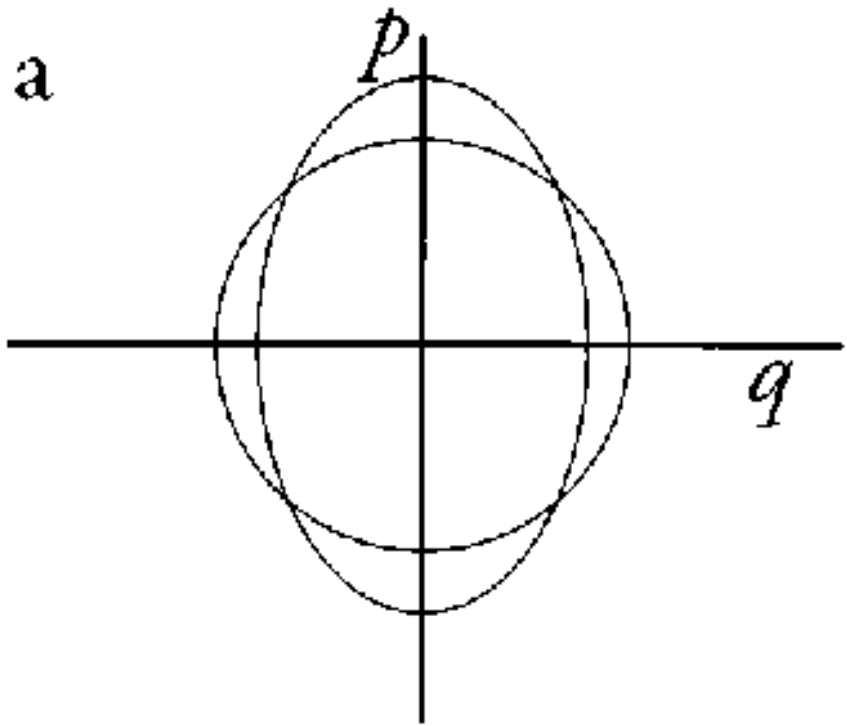
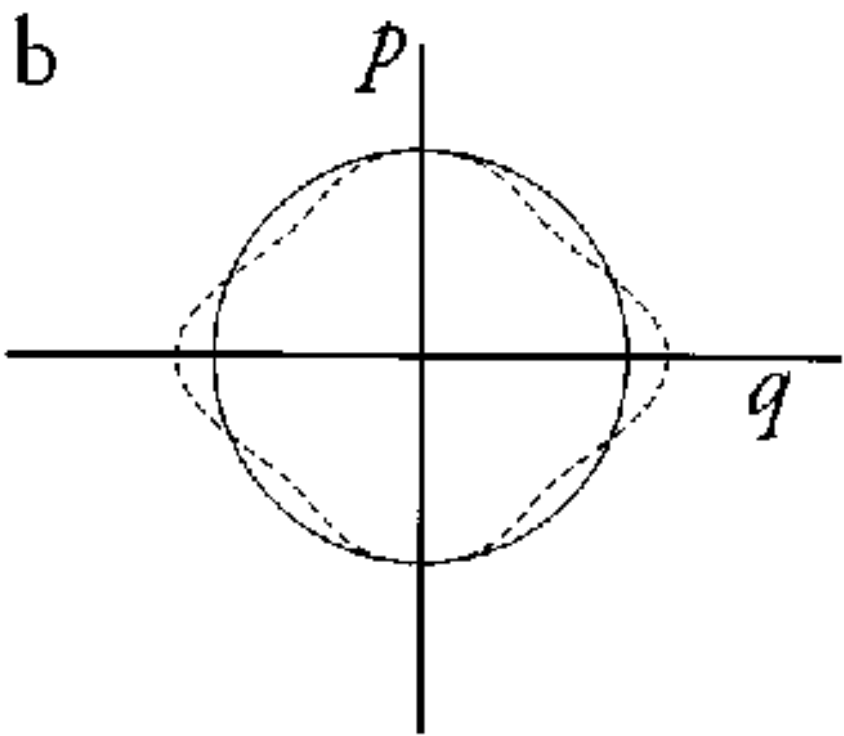
\end{document}